\documentclass[aps,prb,amsmath,amssymb,preprint]{revtex4-1}
\usepackage[utf8]{inputenc} 

\usepackage{geometry} 
\usepackage{graphicx} 
\usepackage{color}

\usepackage{booktabs} 
\usepackage{array} 
\usepackage{paralist} 
\usepackage{verbatim} 
\usepackage{subfig} 
\usepackage{hyperref}

\newcommand*{\citen}[1]{%
  \begingroup
    \romannumeral-`\x 
    \setcitestyle{numbers}%
    \cite{#1}%
  \endgroup   
}

\makeatletter
\newcommand*{\rom}[1]{\expandafter\@slowromancap\romannumeral #1@}
\makeatother

\def\bra#1{\langle#1| }
\def\ket#1{| #1 \rangle }

\begin{document}

\title{The universality of electronic friction II: Equivalence of the quantum-classical Liouville equation approach with von Oppen's nonequilibrium Green's function methods out of equilibrium}
\author{Wenjie Dou and Joseph E. Subotnik}
\affiliation{Department of Chemistry, University of Pennsylvania, Philadelphia, Pennsylvania 19104, USA }

\begin{abstract}

In a recent publication [W. Dou, G. Miao, and J. E. Subotnik, Phys. Rev. Lett. \textbf{119}, 046001 (2017)], using the quantum-classical Liouville equation (QCLE), we derived a very general form for the electronic friction felt by a molecule moving near one or many metal surfaces. Moreover, we have already proved the equivalence of the QCLE electronic friction with the Head-Gordon--Tully model as well as a generalized version of von Oppen's nonequilibrium Green's function (NEGF) method \textit{at equilibrium}  [W. Dou and J. E. Subotnik, Phys. Rev. B \textbf{96}, 104305 (2017)]. In the present paper, we now further prove the equivalence between the QCLE friction and the NEGF friction for the case of multiple metal surfaces and an out-of-equilibrium electronic current. The present results conclude our recent claim that there is only one \textit{universal} electronic friction tensor arising from the Born-Oppenheimer approximation.

\end{abstract}

\maketitle


\section{Introduction}

The dynamics of molecules at molecule-metal interfaces often go beyond the Born-Oppenheimer approximation, where the interplay of electron and nuclei can give rise to a host of nonadiabatic effects. \cite{wodtkereview2004,Saalfrank2013,liu2017energy,chen2016anderson,kisiel2011suppression,ThossHEOM,prldata,zhu2002electron,ibach2013electron,flynn1996vibrational,bijayprb2015,MishaJPCL2015,li2002monte,rittmeyer2018energy,PhysRevLett.118.256001,PhysRevLett.100.116102} These nonadiabatic effects can be seen in many systems.  For the case of a single metal surface at equilibrium, a simple scattering process can reveal unexpected vibrational or translational kinetic energy losses for the molecule due to electronic excitations in the metal as induced by nuclear movement.\cite{kruger2016vibrational, science2000, bunermann2015electron, kruger2016vibrational,shenvi:2009:science} For the case of two or more metal surfaces out of equilibrium, e.g., a molecular junction, under an applied voltage bias with an electronic current running through the molecule, non-Born-Oppenheimer forces can result in heating, \cite{volkovich2011bias, heating,prbqme,lvPRBfriction} photo (or current) induced chemistry, \cite{zang2017electronically, mohn2010reversible, jia2016covalently, jorn2010implications} Franck-Condon blockades, \cite{koch2005franck,hsu2010transport, burzuri2014franck} switching, \cite{Galperinnano,lortscher2006reversible, mishaScience2008,bistabilityPRB2013,paperVI} instability, \cite{thossinstabilities, runaway, prbnitzan} or pumping of the molecule. \cite{beilstein, PhysRevB.95.155431,PhysRevA.71.012331}

Over the past several decades, in order 
to describe such nonadiabatic effects at molecule-metal interface, many researchers have adopted the idea of ``electronic friction", such that the nuclei move on a single potential of mean force, while experiencing a frictional force and a random force induced by electronic motion. \cite{FrictionTully, PERSSON1980175, paperIV} In the literature, quite a few forms of electronic friction have been derived, using a variety of methodologies, \cite{brandbyge, vonOppenPRB, mulFrictionPaperJCP2016} with or without electron-electron (el-el) interactions, \cite{LangrethPRB1998, PhysRevB.11.2122,dou2017born} including or not including non-Condon effects, \cite{Mizielinski2005, dou2017electronic, brandbyge, mishaPRBfriction} invoking a Markovian or non-Markovian frictional kernel, \cite{smith1993electronic,ryabinkin2016mixed,PhysRevLett.49.662,dou2017born} and addressing equilibrium or nonequilibrium scenarios (or both). \cite{Mozyrsky, daligault2007ion,beilstein,dou2017born} Electronic friction has been widely and successfully used to treat vibrational relaxation, chemisorption, and photo induced reaction \textit{et al} at molecule-metal interfaces. \cite{tullyfrictionexpPRL1996,tullyfrictionreview,PhysRevLett.116.217601,PhysRevLett.118.256001,luntz2006femtosecond}

Now,  in the equilibrium scenario, probably the most commonly used electronic friction was given by Head-Gordon and Tully (HGT) in 1995. \cite{FrictionTully} Starting from Ehrenfest dynamics, the HGT friction was derived at zero temperature assuming no el-el interactions, and later extended to finite temperature by ansatz. \cite{head1992vibrational, PhysRevLett.116.217601} At finite temperature, the second fluctuation-dissipation theorem was assumed rather than derived, i.e. a fluctuating force was added by hand to guarantee the nuclear degrees of freedom (DoFs) reach the same temperature with the electronic temperature. 
Besides Head-Gordon and Tully, many other researchers have also derived similar electronic friction tensors,  including, for example, Suhl (Eq. 27 in Ref. \citen{PhysRevB.11.2122}), Galperin (Eq. 17 in Ref. \citen{mishaPRBfriction}), Brandbyge (Eq. A49 in Ref. \citen{brandbyge}), Persson (Eq. 32 in Ref. \citen{Mizielinski2005}), and Hynes (non-Markovian kernel, Eq. 3.18 in Ref. \citen{smith1993electronic}). Of the list above, perhaps the most important contributions have been from Suhl (who first extrapolated that electronic friction should look like a force-force correlation function). \cite{PhysRevB.11.2122} A further advance was made by Daligault and Mozyrsky who derived the random force for the HGT model at equilibrium for finite temperature. \cite{daligault2007ion}

As far as the   
nonequilibrium scenario is concerned,  the situation becomes more complicated and there has been far less development. To our knowledge, the most general nonequilibrium, Markovian electronic friction tensor was given by von Oppen and coworkers, using a nonequilibrium Green's function (NEGF) and a scattering matrix formalism. \cite{beilstein}
Quite different from the equilibrium case, where the electronic friction is a simple damping force, i.e. positive definite and symmetric along nuclear DoFs, the \textit{non}equilibrium electronic friction is no longer symmetric, and can be even negative. Furthermore, the second fluctuation-dissipation theorem breaks down, where the electronic current leads to the heating or pumping of the molecule. 
The von Oppen result should hold for small nuclear velocities assuming that there are no el-el interactions and that there are no non-Condon effects.

At this point in time, given the plethora of different  results discussed above, one of our ongoing research goals has been to compare and connect different approaches for electronic friction and ascertain whether an unifying form exists.  And in fact, recently, in Ref. \citen{dou2017born}, we successfully derived a universal electronic friction from a quantum-classical Liouville equation (QCLE), 
that should be valid in and out of equilibrium, with or without el-el interactions: 
\begin{eqnarray} \label{eq:ManyBodyFriction}
\gamma_{\mu \nu}(\bold R) = - \int_0^{\infty} dt \:  tr_e \left(\partial_\mu \hat H (\bold R) e^{-i \hat H(\bold R) t /\hbar    } \partial_\nu \hat \rho_{ss}(\bold R) e^{i \hat H(\bold R) t  /\hbar  } \right) ,
\end{eqnarray}  
where $\mu$ and $\nu$ are nuclear DoFs, and $\hat H (\bold R)$ is the electronic Hamiltonian, $\hat \rho_{ss}(\bold R)$ is the steady states electronic density matrix. $tr_e$ implies tracing over many-body electronic states. Thus far, we have shown (i) that Eq. \ref{eq:ManyBodyFriction} reduces to Suhl's results as well as the HGT model at equilibrium (without el-el interactions). Furthermore (ii), Ref. \citen{dou2017electronic} shows that, at equilibrium, the HGT model is comparable with the results of Brandbyge and Galperin. Moreover (iii), Ref. \citen{PhysRevB.96.104305} demonstrates that the generalized NEGF electronic friction agrees with HGT model at equilibrium. Thus, altogether,  we have been able to connect the QCLE, NEGF, and HGT friction at equilibrium. Finally, we have also shown that the non-Morkovian friction suggested by Hynes has a natural QCLE expression. \cite{smith1993electronic, dou2017born}

In all of the comparative work above, however, one essential element has been missing. While we have proven that all of the Markovian results of Mozyrsky, Suhl, Persson, Galperin, Hynes, Brandbyge, HGT, and von Oppen reduce to Eq. \ref{eq:ManyBodyFriction} at equilibrium, no further consideration has yet been established for the out of equilibrium scenario. Thus,
in this article, we will take one step further, and prove that, without el-el interactions, indeed the QCLE friction (Eq. \ref{eq:ManyBodyFriction}) reduces to von Oppen's generalized NEGF friction in case of two metals \textit{out of equilibrium}. This agreement greatly strengthens our claim that there is only one, unique electronic friction associated with the Born-Oppenheimer approximation. In what follows, we will also provide an explicit, very general formula for calculating that friction tensor in the limit of no el-el interactions; our work will include  
 non-Condon effects and thus go beyond von Oppen's results.

We organize the structure of the paper as follows. In Sec. \ref{sec:theory}, we explain our model and provide important relationships that will be used later on. In Sec. \ref{sec:vonOppen}, we demonstrate the agreement between QCLE friction and NEGF friction. In Sec. \ref{sec:nonCondon}, we adopt the commonly used molecule-metal Hamiltonian and evaluate the nonequilibrium electronic friction tensor while accounting for non-Condon effects. We conclude in Sec. \ref{sec:con}.  

Regarding the notation, we use $p$ and $q$ to denote electronic orbitals in general, $m$ and $n$ for the electronic orbitals in a molecule (dots), and $k$ and $k'$ for the electronic orbitals in a metal (lead). We further use $\alpha=L,R$ to signify the left or right metal.  $\mathcal{G}$ will denote the total system (dots plus leads) steady-state non-equilibrium Green's functions, and $G$ will denote the dots' (i.e. molecules') steady-state non-equilibrium Green's functions. We use $\mu$ (or $\nu$) to denote nuclear degrees of freedom (DoFs), and we use $\mu_L$ (and $\mu_R$) to denote the Fermi level of the left (and right) metal.


\section{quadratic Hamiltonian}   \label{sec:theory}

We consider 
a total Hamiltonian $\hat H_{tot}$ which can be  divided into an electronic Hamiltonian $\hat H$ and a nuclear kinetic energy operator: 
\begin{eqnarray} 
\hat H_{tot}= \hat H  +  \sum_{\mu} \frac{P_{\mu}^2} { 2m^{\mu} }.
\end{eqnarray}
The electronic Hamiltonian $\hat H$ consists of a manifold of electrons that is quadratic (in electronic orbitals $p$, $q$) plus a pure nuclear potential energy $U_0(\bold R)$:  
\begin{eqnarray} \label{eq:quadH}
\hat H =  \sum_{pq}  \mathcal{H}_{pq} (\bold R)  \hat d^\dagger_p \hat d_q + U_0 (\bold R) .
\end{eqnarray}

For such an electronic Hamiltonian (Eq. \ref{eq:quadH}, without el-el interaction), the general form of the electronic friction (Eq. \ref{eq:ManyBodyFriction}) can be recast into the single particle basis (as shown in Appendix \ref{app:a}), 
\begin{eqnarray} \label{eq:frictionSP}
\gamma_{\mu\nu} = - \hbar \int  \frac{d\epsilon}{2\pi} \:  Tr_m \left( \partial_\mu \mathcal{H}  \mathcal{G}^R(\epsilon)  \partial_\nu  \sigma_{ss}  \mathcal{G}^A (\epsilon) \right) . 
\end{eqnarray}  
Here 
\begin{eqnarray}
\mathcal{G}^{R/A} (\epsilon) = \frac{1}{ \epsilon -  \mathcal{H}  \pm  i\eta} 
\end{eqnarray} 
are retarded and advanced Green's function of the electrons respectively ($\eta$ is an positive infinitesimal). Thus, for the NEGFs, one can easily establish the following identities, 
\begin{eqnarray} \label{eq:identities1}
 \partial_\nu \mathcal{G}^R (\epsilon') = \mathcal{G}^R (\epsilon')  \partial_\nu \mathcal{H} \mathcal{G}^R (\epsilon'),  \: \partial_{\epsilon'} \mathcal{G}^R (\epsilon') = - \mathcal{G}^R (\epsilon') \mathcal{G}^R (\epsilon'), \\
  \partial_\nu \mathcal{G}^A (\epsilon') = \mathcal{G}^A (\epsilon')  \partial_\nu \mathcal{H} \mathcal{G}^A (\epsilon'),  \: \partial_{\epsilon'} \mathcal{G}^A (\epsilon') = - \mathcal{G}^A (\epsilon') \mathcal{G}^A (\epsilon') \label{eq:identities2}.
\end{eqnarray} 

Besides the retarded and advanced GFs, we also find in Eq. \ref{eq:frictionSP}  the steady-state electronic population matrix $\sigma^{ss}_{qp} = tr_e(\hat \rho_{ss} \hat d^\dagger_p \hat d_q)$. 
$\sigma_{ss}$ is usually  expressed in terms of the lesser NEGF  $\mathcal{G}^<$, 
\begin{eqnarray} \label{eq:sigmaG}
\sigma_{ss} =   \int \frac{d\epsilon'}{2\pi i } \mathcal{G}^< (\epsilon')  ,
\end{eqnarray}
where $\mathcal{G}^<(\epsilon')$ is the Fourier transform of $\mathcal{G}^< (t_1, t_2)$. 
In turn, the lesser nonequilibrium Green's function (NEGF) $\mathcal{G}^< (t_1, t_2)$ is defined as
\begin{eqnarray} 
 \mathcal{G}^<_{qp} (t_1, t_2) &=& \frac{i}{\hbar} tr_e(\hat \rho_{ss}  \hat d^\dagger_p (t_2) \hat d_q (t_1) ) ,
\end{eqnarray}
where the electronic operators are written in the Heisenberg picture 
$\hat d^\dagger_p (t)  \equiv e^{i \hat H t /\hbar } \hat d^\dagger_p e^{ -i \hat H t /\hbar } $
(and $\hat d_q (t)  \equiv e^{i \hat H t /\hbar } \hat d_q e^{ -i \hat H t /\hbar }  $). Note that, in the single particle basis, $Tr_m$ implies summing over the electronic orbitals ($p$ and $q$); vice versa, in the many-particle basis, $tr_e$ implies a trace of all many-body electronic states.

In order to evaluate $\sigma_{ss}$ in Eq.  \ref{eq:sigmaG}, we must first evaluate the lesser NEGF  $\mathcal{G}^<$ and the derivative $\partial_\nu  \mathcal{G}^< $. To do so, 
 we invoke the Keldysh equation, 
\begin{eqnarray} \label{eq:LangrethRule}
\mathcal{G}^< (\epsilon')  = \mathcal{G}^R (\epsilon')   \Pi^< \mathcal{G}^A (\epsilon') , 
\end{eqnarray}
where $\Pi^<$ is the total electronic lesser self-energy.  

Below we will adopt a dot-lead (system-bath) separation (Eqs. \ref{eq:ah1}-\ref{eq:ah4}), such that $\Pi^<$ can be written explicitly (Eq. \ref{eq:PiLesser} in Appendix \ref{app:b}).    
That being said, we emphasize that all of the results below do not depend on the exact value of $\Pi^<$. We require only that $\Pi^<$ does not depend on energy ($\epsilon'$) or position ($\bold R$). As outlined in Eq. \ref{eq:PiLesser}, these assumptions about $\Pi^<$ follow because the bath Hamiltonian does not depend on $\bold R$. 

Since $\Pi^<$ does not depend on position ($\bold R$), together with the Keldysh equation (Eq. \ref{eq:LangrethRule}) and the identities in Eqs. \ref{eq:identities1}-\ref{eq:identities2}, it is straightforward to show that 
\begin{eqnarray} \label{eq:partialG}
\partial_\nu  \mathcal{G}^<   &=&  \mathcal{G}^R  \partial_\nu \mathcal{H}  \mathcal{G}^R  \Pi^< \mathcal{G}^A   +  \mathcal{G}^R   \Pi^< \mathcal{G}^A   \partial_\nu \mathcal{H}  \mathcal{G}^A  \nonumber \\
&=& \mathcal{G}^R  \partial_\nu \mathcal{H}  \mathcal{G}^< +  \mathcal{G}^<   \partial_\nu \mathcal{H}  \mathcal{G}^A  .
\end{eqnarray}

With the above identities (Eqs. \ref{eq:LangrethRule}-\ref{eq:partialG}), below we will show that Eq. \ref{eq:frictionSP} reduces to the following NEGF result (derived in Ref. \citen{PhysRevB.96.104305}),
\begin{eqnarray}  \label{eq:frictionBeil}
\gamma_{\mu\nu} 
 =  \hbar \int  \frac{d\epsilon}{2\pi} \:  Tr_m \left( \partial_\mu \mathcal{H}   \partial_\epsilon \mathcal{G}^R(\epsilon)   \partial_\nu \mathcal{H}  \mathcal{G}^< (\epsilon)   
 - \partial_\mu \mathcal{H}  \mathcal{G}^< (\epsilon)   \partial_\nu \mathcal{H}  \partial_\epsilon \mathcal{G}^A(\epsilon)   \right)  .
\end{eqnarray} 


\section{Agreement of QCLE friction and NEGF friction} \label{sec:vonOppen}
To prove the equivalence  between Eq. \ref{eq:frictionSP} and Eq. \ref{eq:frictionBeil}, we use the eigenbasis of the electronic Hamiltonian $\mathcal{H}$, $\mathcal{H} \ket m = \epsilon_m \ket m $, such that  Eq. \ref{eq:frictionSP} can be expressed as 
\begin{eqnarray} \label{eq:FrictionFirst}
\gamma_{\mu\nu}  =&& - \hbar  \sum_{mn} \int  \frac{d\epsilon}{2\pi} \:  \bra n \partial_\mu \mathcal{H} \ket m \frac{1}{\epsilon- \epsilon_m +i\eta}  \bra m \partial_\nu  \sigma_{ss} \ket n   \frac{1}{\epsilon- \epsilon_n  - i\eta}    \nonumber \\
 =&& - i \hbar  \sum_{mn}  \:  \bra n \partial_\mu \mathcal{H} \ket m \frac{1}{\epsilon_n - \epsilon_m +2i\eta}  \bra m \partial_\nu  \sigma_{ss} \ket n     .
\end{eqnarray}  
In the above equation, we have used the residue theorem for contour integration.  Using  Eq. \ref{eq:sigmaG} and Eq. \ref{eq:partialG}, $\bra m \partial_\nu  \sigma_{ss} \ket n$ in Eq. \ref{eq:FrictionFirst} can be rewritten as 
\begin{eqnarray} \label{eq:partialSigma}
\bra m \partial_\nu  \sigma_{ss} \ket n &=&   \int \frac{d\epsilon'}{2\pi i } \bra m \partial_\nu \mathcal{G}^< (\epsilon')  \ket n  \nonumber \\
&=&  \int \frac{d\epsilon'}{2\pi i } \bra m \mathcal{G}^R(\epsilon')  \partial_\nu \mathcal{H}  \mathcal{G}^<(\epsilon') +  \mathcal{G}^< (\epsilon')  \partial_\nu \mathcal{H}  \mathcal{G}^A (\epsilon')  \ket n .
\end{eqnarray}  

%

Note that the second term in Eq.  \ref{eq:partialSigma} is the Hermitian conjugate of the first term. 
We now evaluate the first term of Eq. \ref{eq:partialSigma}. In the eigenbasis of the electronic Hamiltonian, with the definition of $\mathcal{G}^{R/A}$ and the Keldysh equation (Eq. \ref{eq:LangrethRule}), we have
\begin{eqnarray}
 &&\int \frac{d\epsilon'}{2\pi i } \bra m \mathcal{G}^R(\epsilon')  \partial_\nu \mathcal{H}  \mathcal{G}^< (\epsilon')  \ket n   \nonumber \\
 =&&  \int \frac{d\epsilon'}{2\pi i } \bra m  \mathcal{G}^R (\epsilon')  \partial_\nu \mathcal{H}  \mathcal{G}^R (\epsilon')  \Pi^< \mathcal{G}^A (\epsilon')  \ket n  \nonumber \\
 =&& \sum_{m'}  \int \frac{d\epsilon'}{2\pi i }  \frac{1}{\epsilon'- \epsilon_m +i\eta} \bra m    \partial_\nu \mathcal{H} \ket {m'}  \frac{1}{\epsilon'- \epsilon_{m'} +i\eta} \bra {m'} \Pi^<  \ket n   \frac{1}{\epsilon'- \epsilon_n  - i\eta}     .
\end{eqnarray} 
As stated before, $\Pi^<$ does not depend on energy ($\epsilon'$), such that we can apply the residue theorem to the above equation,  
\begin{eqnarray} \label{eq:SigmaFirst}
 &&\int \frac{d\epsilon'}{2\pi i } \bra m \mathcal{G}^R(\epsilon')  \partial_\nu \mathcal{H}  \mathcal{G}^< (\epsilon')  \ket n   \nonumber \\
 =&& \sum_{m'}   \frac{1}{\epsilon_n - \epsilon_m + 2i\eta} \bra m    \partial_\nu \mathcal{H} \ket {m'}  \frac{1}{\epsilon_n - \epsilon_{m'} + 2 i\eta} \bra {m'} \Pi^<  \ket n    .
\end{eqnarray} 
A similar analysis applies to the second term of Eq. \ref{eq:partialSigma}. 

At this point, we consider the first term of Eq. \ref{eq:frictionBeil}. Again, using the definition of $\mathcal{G}^{R/A}$ as well as the Keldysh equation, and applying the residue theorem, we find
\begin{eqnarray} 
 &&-  \int  \frac{d\epsilon}{2\pi} \:  Tr_m \left( \partial_\mu \mathcal{H}   \partial_\epsilon \mathcal{G}^R(\epsilon)   \partial_\nu \mathcal{H}  \mathcal{G}^< (\epsilon)   \right)  \nonumber \\
= && \int  \frac{d\epsilon}{2\pi} \:  Tr_m \left( \partial_\mu \mathcal{H}  \mathcal{G}^R(\epsilon)   \mathcal{G}^R(\epsilon)  \partial_\nu \mathcal{H}  \mathcal{G}^R (\epsilon)  \Pi^< \mathcal{G}^A (\epsilon) \right)  \nonumber \\
= &&  \sum_{mnm'} \int  \frac{d\epsilon}{2\pi} \:  \bra n \partial_\mu \mathcal{H} \ket m \frac{1}{(\epsilon- \epsilon_m +i\eta)^2} \bra m \partial_\nu \mathcal{H} \ket {m'}  \frac{1}{\epsilon- \epsilon_{m'} +i\eta}  \bra {m'} \Pi^<   \ket {n}    \frac{1}{\epsilon- \epsilon_n  - i\eta}  \nonumber \\
= && i  \sum_{mnm'}  \:  \bra n \partial_\mu \mathcal{H} \ket m \frac{1}{(\epsilon_n- \epsilon_m +2i\eta)^2}   \bra m \partial_\nu \mathcal{H} \ket {m'}  \frac{1}{\epsilon_n- \epsilon_{m'} +2 i\eta}  \bra {m'} \Pi^<   \ket {n}   .
\end{eqnarray} 
Comparing the above equation with Eq. \ref{eq:SigmaFirst}, we have the following identity:  
\begin{eqnarray}  \label{eq:part1}
 &&  \int  \frac{d\epsilon}{2\pi} \:  Tr_m \left( \partial_\mu \mathcal{H}   \partial_\epsilon \mathcal{G}^R(\epsilon)   \partial_\nu \mathcal{H}  \mathcal{G}^< (\epsilon)   \right)  \nonumber \\
 = && - i  \sum_{mn}  \:  \bra n \partial_\mu \mathcal{H} \ket m \frac{1}{\epsilon_n - \epsilon_m +2i\eta}  \int \frac{d\epsilon'}{2\pi i } \bra m \mathcal{G}^R(\epsilon')  \partial_\nu \mathcal{H}  \mathcal{G}^< (\epsilon')  \ket n    .
\end{eqnarray}

Similarly, we can show
\begin{eqnarray}  \label{eq:part2}
 && - \int  \frac{d\epsilon}{2\pi} \:  Tr_m \left( \partial_\mu \mathcal{H}  \mathcal{G}^< (\epsilon)   \partial_\nu \mathcal{H}  \partial_\epsilon \mathcal{G}^A(\epsilon)   \right)  \nonumber \\
 = &&-  i   \sum_{mn}  \:  \bra n \partial_\mu \mathcal{H} \ket m \frac{1}{\epsilon_n - \epsilon_m +2i\eta}  \int \frac{d\epsilon'}{2\pi i } \bra m \mathcal{G}^< (\epsilon')  \partial_\nu \mathcal{H}  \mathcal{G}^A (\epsilon')  \ket n    .
\end{eqnarray} 
Note that Eq. \ref{eq:part2} is the Hermitian conjugate of Eq. \ref{eq:part1}. 

Finally, if we put Eq. \ref{eq:partialSigma} back into Eq. \ref{eq:FrictionFirst}, together with the relationships shown in Eq. \ref{eq:part1} and Eq. \ref{eq:part2}, we recover
\begin{eqnarray} 
\gamma_{\mu\nu} =&& - \hbar \int  \frac{d\epsilon}{2\pi} \:  Tr_m \left( \partial_\mu \mathcal{H}  \mathcal{G}^R(\epsilon)  \partial_\nu  \sigma_{ss}  \mathcal{G}^A (\epsilon) \right) \nonumber \\
 = && \hbar \int  \frac{d\epsilon}{2\pi} \:  Tr_m \left( \partial_\mu \mathcal{H}   \partial_\epsilon \mathcal{G}^R(\epsilon)   \partial_\nu \mathcal{H}  \mathcal{G}^< (\epsilon)   
 - \partial_\mu \mathcal{H}  \mathcal{G}^< (\epsilon)   \partial_\nu \mathcal{H}  \partial_\epsilon \mathcal{G}^A(\epsilon)   \right)  .
\end{eqnarray} 
Thus, we have proven our claim that QCLE friction (Eq. \ref{eq:frictionSP}) agrees with NEGF friction (Eq. \ref{eq:frictionBeil}). Note that Eq. \ref{eq:ManyBodyFriction} is much more general than any of these expressions since el-el interactions are allowed in Eq. \ref{eq:ManyBodyFriction}, whereas el-el interactions are absent from Eq. \ref{eq:frictionSP} as well as Eq. \ref{eq:frictionBeil}. 


\section{system-bath separation and Non-Condon effects} \label{sec:nonCondon}  
The results in Eq. \ref{eq:frictionSP} are very general and are applicable for any quadratic Hamiltonian without el-el interactions. To investigate non-Condon effects, we now adopt the standard dot-lead separation, such that the total electronic Hamiltonian $\hat H$ can be divided into system $\hat H_s$ and bath $\hat H_b$, as well as system-bath coupling $\hat H_c$,  
\begin{eqnarray} \label{eq:ah1}
 \hat{H} &=& \hat{H}_s+  \hat{H}_b+  \hat{H}_c,  \\
\label{Hs}
 \hat{H}_s &=& \sum_{mn} h_{mn} (\bold R)  \hat{b}^\dagger_m \hat{b}_n + U_0(\bold R), \\
 \hat{H}_b &=& \sum_{k\alpha}  \epsilon_{k\alpha}  \hat{c}_{k\alpha}^\dagger  \hat{c}_{k\alpha}, \label{eq:ah3} \\
 \label{eq:ah4}
 \hat{H}_c &=& \sum_{m, k \alpha } V_{m, k\alpha} (\bold R)  \hat{b}^\dagger_m  \hat{c}_{k\alpha} +  V_{k\alpha, m} (\bold R) \hat{c}_{k\alpha}^\dagger  \hat{b}_m . 
\end{eqnarray}
Here $m$, $n$ are orbitals in the molecule, and $\alpha=L, R$ indicates left and right leads, which linearly couple to the molecule through $\hat H_c$. We remind the reader that the total Hamiltonian $\hat H_{tot}$ still is a combination of the electronic Hamiltonian $\hat{H}$ with the nuclear kinetic energy,  $\hat H_{tot}= \hat H  +  \sum_{\alpha} \frac{P_{\alpha}^2} { 2m^{\alpha} }$. Note also that the molecule-leads interactions $V_{m, k\alpha} (\bold R) $ also depend on nuclear position $\bold R$, which will give rise to non-Condon effects. 

To evaluate the electronic friction (Eq. \ref{eq:frictionBeil}) and connect to the results in Ref. \citen{beilstein}, we first consider the case where $V_{m, k\alpha} (\bold R)$ does not depend on $\bold R$. In such a case, only $\hat{H}_s$ depends on $\bold R$, and therefore:  
\begin{eqnarray}
Tr_m \left( \partial_\mu \mathcal{H}   \partial_\epsilon \mathcal{G}^R  \partial_\nu \mathcal{H}  \mathcal{G}^<  \right)
= &&  \sum_{mnm'n'} \partial_\mu h_{mn} \partial_\epsilon G^R_{nm'}   \partial_\nu h_{m'n'}  G^<_{n'm}  \nonumber \\
= && Tr_s \left( \partial_\mu h \partial_\epsilon G^R \partial_\nu h G^< \right) .
\end{eqnarray}
Here $Tr_s$ implies summation over system orbitals ($m$ and $n$), and $G^R = (\epsilon-h-\Sigma^R)^{-1}$ is the system retarded GF. $\Sigma^R_{mn}=\sum_{k\alpha} V_{m,k\alpha} g^R_{k\alpha} V_{k\alpha,n} $
is the system retarded self-energy; $G^<$ is the system lesser GF (see Appendix \ref{app:b}). Thus, 
without any non-Condon contributions, the nonequilibrium electronic friction is 
\begin{eqnarray}  \label{eq:vonOppenResult}
\gamma_{\mu\nu} =  \hbar \int  \frac{d\epsilon}{2\pi} \: &&Tr_s \left( \partial_\mu h \partial_\epsilon G^R 
\partial_\nu h G^<  \right) + \mbox{h.c.}, 
\end{eqnarray}
which reduces to von Oppen's results in Ref. \citen{beilstein}. Here, $\mbox{h.c.}$ denotes the Hermitian conjugate. 

Second, for the case where $V_{m, k\alpha} (\bold R)$ does depend on $\bold R$, the results are much more complicated. However, in the wide-band approximation, as shown in the supplemental material (SM), the result can be simplified as 
\begin{eqnarray}
&&Tr_m \left( \partial_\mu \mathcal{H}   \partial_\epsilon \mathcal{G}^R  \partial_\nu \mathcal{H}  \mathcal{G}^<  \right) \nonumber \\
=&& Tr_s \left( (\partial_\mu h \partial_\epsilon G^R  +  \bar \Sigma_{\mu}^R \partial_\epsilon G^R +  \tilde \Sigma^A_{\mu} \partial_\epsilon G^R ) 
(\partial_\nu h G^<  +  \partial_\nu  \Sigma^R  G^< +  \bar \Sigma^<_{\nu} G^A ) \right) \nonumber \\
+ &&Tr_s(  \tilde \Sigma^<_{\mu} \partial_\epsilon G^R  (\partial_\nu h + \partial_\nu \Sigma^R)G^R+   \partial_\epsilon G^R  \Sigma^<_{\nu, \mu}  ) .
\end{eqnarray}
Again, $Tr_s$ implies summation over system orbitals ($m$ and $n$). We have further defined the following quantities, 
\begin{eqnarray} \label{eq:Sigma1}
\bar \Sigma^R_{\mu, mn} &=&\sum_{k\alpha} \partial_\mu V_{m,k\alpha} g^r_{k\alpha} V_{k\alpha,n} ,\\
\tilde \Sigma^A_{\mu, mn}&=&\sum_{k\alpha}  V_{m,k\alpha} g^a_{k\alpha} \partial_\mu V_{k\alpha,n},\\
\tilde \Sigma^<_{\mu, mn}&=&\sum_{k\alpha}  V_{m,k\alpha} g^<_{k\alpha} \partial_\mu V_{k\alpha,n} ,\\
\bar \Sigma^<_{\mu, mn}&=&\sum_{k\alpha}  \partial_\mu V_{m,k\alpha} g^<_{k\alpha}  V_{k\alpha,n} ,\\
 \Sigma^<_{\nu\mu, mn}&=&\sum_{k\alpha}  \partial_\nu V_{m,k\alpha} g^<_{k\alpha}   \partial_\mu V_{k\alpha,n} .\label{eq:Sigma5}
\end{eqnarray}
$g^r_{k\alpha}$, $g^a_{k\alpha}$, and $g^<_{k\alpha}$ are the zero order retarded, advanced, and lesser Green's functions respectively for the non-interacting leads. The explicit forms are given in Appendix \ref{app:b}.  
With these definitions, the nonequilibrium and non-Condon electronic friction can be written as 
\begin{eqnarray}  \label{eq:nonCondonFriction}
\gamma_{\mu\nu} =  \hbar \int  \frac{d\epsilon}{2\pi} \: &&Tr_s \left( (\partial_\mu h \partial_\epsilon G^R  +  \bar \Sigma_{\mu}^R \partial_\epsilon G^R +  \tilde \Sigma^A_{\mu} \partial_\epsilon G^R ) 
(\partial_\nu h G^<  +  \partial_\nu  \Sigma^R  G^< +  \bar \Sigma^<_{\nu} G^A ) \right) \nonumber \\
+ &&Tr_s(  \tilde \Sigma^<_{\mu} \partial_\epsilon G^R  (\partial_\nu h + \partial_\nu \Sigma^R)G^R+   \partial_\epsilon G^R  \Sigma^<_{\nu \mu}  ) + h.c.
\end{eqnarray}
Again, in the Condon approximation, Eqs. \ref{eq:Sigma1}-\ref{eq:Sigma5} vanish, such that the above equation (Eq. \ref{eq:nonCondonFriction}) reduces to von Oppen's result (Eq. \ref{eq:vonOppenResult}). 

\subsection{A single level with a harmonic oscillator}
We will now apply the results above to the case of a single level (i.e. a dot) coupled to a harmonic oscillator (nuclear DoF) and two metallic baths. The corresponding system Hamiltonian is  
\begin{eqnarray}
\hat{H}_s =  \epsilon_b (x) \hat{b}^\dagger  \hat{b} + \frac12 m\omega^2 x^2 , 
\end{eqnarray} 
where we assume $\epsilon_b(x)$ depends linearly on $x$:
\begin{eqnarray}
\epsilon_b(x) = \epsilon_0+ \lambda x \sqrt{m\omega/\hbar} . 
\end{eqnarray} 
 The single level is coupled to the left and right leads through the following Hamiltonian: 
 \begin{eqnarray}
 \hat{H}_c = \sum_{k \alpha} V_{k\alpha} (x)  ( \hat{b}^\dagger  \hat{c}_{k\alpha} +  \hat{c}_{k\alpha}^\dagger  \hat{b} ). 
 \end{eqnarray} 
Below we will apply the wide-band approximation, such that $V_{k\alpha} (x)$ is independent of $k$. We take $V_{k\alpha} (x)$ to have the following form (as a function of $x$): 
 \begin{eqnarray} \label{eq:Zx}
 V_{k\alpha} (x) = V_{\alpha} \sqrt{1+ z \exp(- m\omega x^2/\hbar) }\equiv V_{\alpha} Z(x).
 \end{eqnarray} 
Note that if we take $z=0$ in the above equation, $V_{k\alpha} (x)$ will be independent of $x$, i.e. $V_{k\alpha} (x)$ will satisfy the Condon approximation. 

According to the wide-band approximation, the self-energy is purely imaginary and can be defined as
\begin{eqnarray} 
\sum_{k\alpha}  V^2_{k\alpha} (x) g^r_{k\alpha} &=&  \sum_{k\alpha} V_\alpha^2 g^r_{k\alpha} Z^2(x)= - \frac i 2  ( \Gamma_0^L + \Gamma_0^R ) Z^2(x) , \\
\sum_{k\alpha} V^2_{k\alpha} (x)  g^<_{k\alpha}&=& \sum_{k\alpha} V_\alpha^2 g^<_{k\alpha} Z^2(x) =  i   ( \Gamma_0^L  f^L + \Gamma_0^R  f^R ) Z^2(x). 
 \end{eqnarray} 

Now we evaluate Eqs. \ref{eq:Sigma1}-\ref{eq:Sigma5}, using the fact that all $x$ dependence in Eqs. \ref{eq:Sigma1}-\ref{eq:Sigma5} is through the term $Z(x)$ defined in Eq. \ref{eq:Zx}. We sum up all of the relevant terms and calculate the electronic friction according to Eq. \ref{eq:nonCondonFriction}.  

\begin{figure}[htbp] 
   \centering 
     \includegraphics[width=8cm]{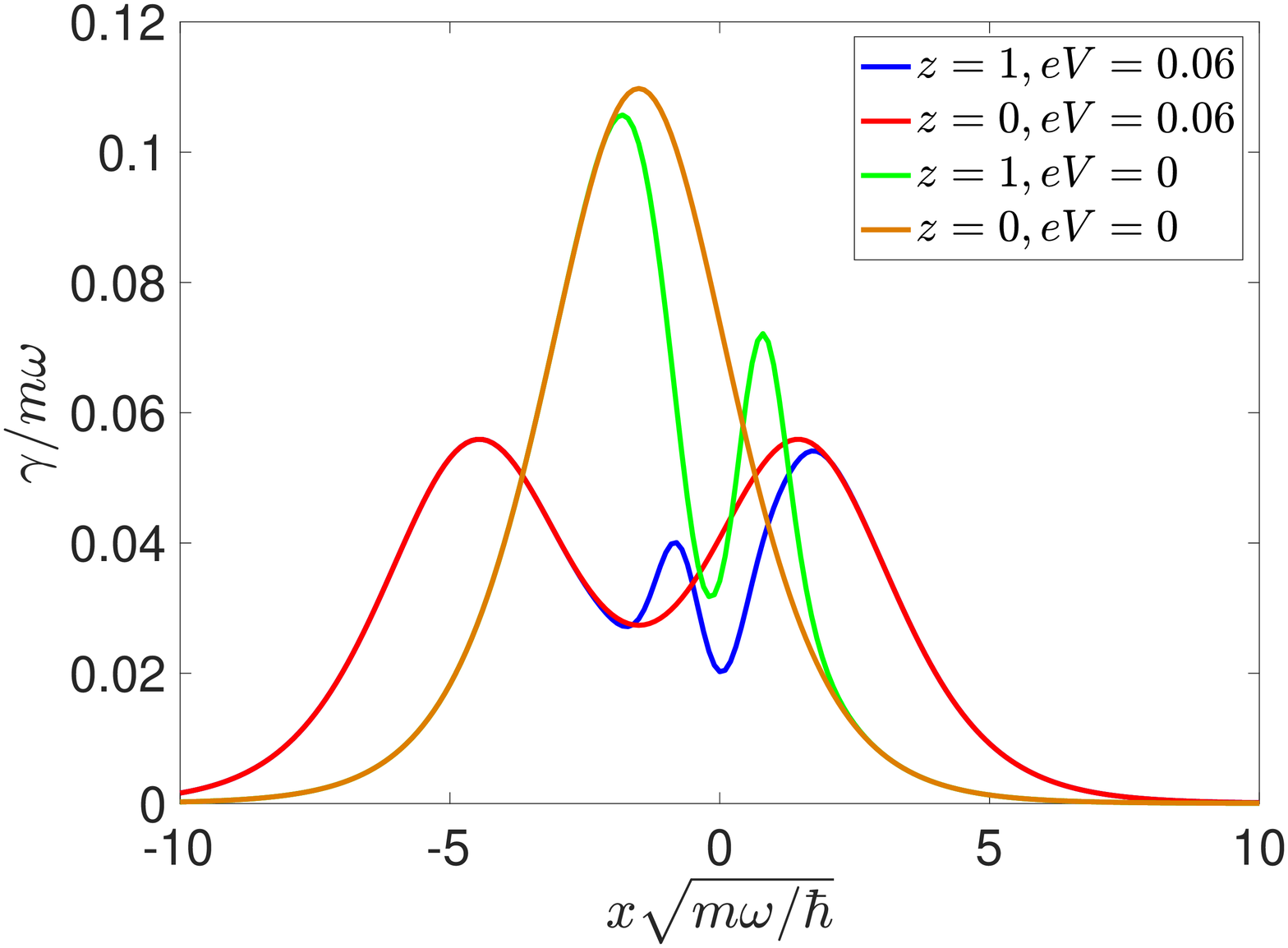}
   \caption{Electronic friction as a function of $x$ for a single level coupled linearly to a harmonic oscillator. For the nonequilibrium case ($eV\ne 0$), when the Condon approximation holds ($z=0$), the electronic friction exhibits two peaks corresponding to the resonance of the dot level with each of the two different Fermi levels for the leads: $\epsilon_b(x) = \mu_L$ and $\epsilon_b(x)=\mu_R$. With non-Condon effects ($z=1$), the electronic friction exhibits a dip at the position $x=0$, where $ V_{k\alpha} (x)$ is maximum.  Thus, when we go out of equilibrium and break the Condon approximation, we effectively find three peaks. $kT=0.01$, $\lambda=0.01$, $\Gamma_0^L=\Gamma_0^R=0.01$, $\hbar\omega=0.003$, $\epsilon_0=0.015$, $\mu_L=-\mu_R=eV/2$. }
   \label{fig:nonCondon}
\end{figure}

In Fig. \ref{fig:nonCondon}, we plot the electronic friction as a function of $x$. For the equilibrium case (i.e. no bias, $eV=0$), when the Condon approximation holds ($z=0$, such that $V_{k\alpha}$ is independent of $x$), the electronic friction exhibits a peak corresponding to the resonance of the dot level with the Fermi level of the leads: $\epsilon_b(x) = \mu_L=\mu_R$. With non-Condon effects ($z=1$), the electronic friction exhibits a dip at the position $x=0$, where $ V_{k\alpha} (x)$ is maximum. 
This change from one peak to effectively two peaks was observed previously (in Ref. \citen{dou2017electronic}) for the equilibrium case of one dot coupled to a single metal lead.

For the nonequilibrium case (i.e. $eV\ne 0$), when the Condon approximation holds ($z=0$), again one peak becomes two peaks, but now for a different reason:  two peaks arise from a  resonance of the dot level with the each of the two different Fermi levels for the left and right leads: $\epsilon_b(x) = \mu_L$ and $\epsilon_b(x) = \mu_R$. Interestingly, when non-Condon effects ($z=1$) are included,  the electronic friction again exhibits a dip at the position where $ V_{k\alpha} (x)$ is maximum, which now effectively  results in three peaks.
Such results demonstrate that, when molecule-metal interactions depend strongly on nuclear geometry, non-Condon effects can strongly influence the relevant nonadiabatic dynamics at metal surfaces.

\section{conclusion} \label{sec:con}
In summary, we have shown that, in the absence of electron-electron interactions, the electronic friction from a quantum-classical Liouville equation (QCLE) reduces to the results from von Oppen's nonequilibrium Green's function (NEGF) method. This agreement holds in general, in or out of equilibrium, for the case of quadratic Hamiltonian. Furthermore, we have shown that non-Condon effects can be easily included into a nonequilibrium electronic friction. 
Thus, given our previous work proving that, at equilibrium, the QCLE friction agrees with the Head-Gordon--Tully model as well as many other forms of electronic friction, \cite{dou2017born, PhysRevB.96.104305, PhysRevB.11.2122, brandbyge,smith1993electronic} we believe there is now very strong proof that, in the limit of Markovian dynamics,  there is only one, \textit{universal} electronic friction associated with the Born-Oppenheimer approximation in the adiabatic limit. Future work must address how to incorporate non-Markovian effects efficiently; is there an optimal approach or many different approaches depending on the Hamiltonian? \cite{lvPRBfriction, dou2017born, smith1993electronic, MishaJPCL2015} 
We will address this question in a future study.

\begin{acknowledgments}

This work was supported by the
(U.S.) Air Force Office of Scientific Research (USAFOSR)
PECASE award under AFOSR Grant No. FA9950-13-1-0157. 

\end{acknowledgments}


\appendix 

\section{friction in the single particle basis} \label{app:a}

The friction tensor in the many-body representation is 
\begin{eqnarray}
\gamma_{\mu \nu} &=& - \int_0^\infty dt \:  tr_e \left( \partial_{\mu} \hat H e^{ - i \hat H t/\hbar } \partial_{\nu} \hat \rho_{ss}  e^{  i \hat H t/\hbar }  \right)  .
\end{eqnarray}
For the quadratic Hamiltonian in Eq. \ref{eq:quadH}, we will recast the above equation into the single particle basis (Eq. \ref{eq:frictionSP}). 

We note first that $U_0 (\bold R)$ does not contribute to the friction, because  
\begin{eqnarray}
 tr_e \left( \partial_{\mu}  U_0  e^{ - i \hat H t /\hbar } \partial_{\nu} \hat \rho_{ss}  e^{  i \hat H t/\hbar }  \right) = \partial_{\mu}  U_0  \:  tr_e \left(  \partial_{\nu} \hat \rho_{ss}  \right) =0. 
\end{eqnarray}
Here, we have used the fact that $tr_e \left( \hat \rho_{ss}  \right) =1$. The friction can be rewritten as  
\begin{eqnarray} \label{eq:friction45}
\gamma_{\mu \nu}
&=& - \int_0^\infty dt \:  tr_e \left( e^{  i \hat H t/\hbar } \partial_{\mu} \hat H e^{ - i \hat H t /\hbar } \partial_{\nu} \hat \rho_{ss}    \right) \nonumber \\
&=& - \int_0^\infty dt \:  \sum_{pq} \partial_{\mu}  \mathcal{H}_{pq}  tr_e \left( e^{  i \hat H t/\hbar } \hat d^\dagger_p \hat d_q e^{ - i \hat H t /\hbar } \partial_{\nu} \hat \rho_{ss}    \right) .
\end{eqnarray}

We proceed to evaluate  
\begin{eqnarray}
\hat d^\dagger_p (t)  = e^{i \hat H t /\hbar } \hat d^\dagger_p e^{ -i \hat H t /\hbar }  ,\\
\hat d_q (t)  = e^{i \hat H t /\hbar } \hat d_q e^{ -i \hat H t /\hbar }  .
\end{eqnarray}
The time derivatives of these operators are 
\begin{eqnarray}
\dot{\hat{d}}^\dagger_p (t)  = e^{i \hat H t}  \frac{i}{\hbar}  [ \hat H,  \hat d^\dagger_p ] e^{- i \hat H t}  =   \frac{i}{\hbar} \sum_{a} \hat d^\dagger_a (t)  \mathcal{H}_{ap} , \\
\dot{\hat{d}}_q (t)  = e^{i \hat H t}  \frac{i}{\hbar}  [\hat H,  \hat d_q ] e^{- i \hat H t}  =  -  \frac{i}{\hbar} \sum_{b} \mathcal{H}_{qb} \hat d_b (t) .
\end{eqnarray}
The above equations can be solved 
\begin{eqnarray}
 e^{i \hat H t} \hat d^\dagger_p  e^{ - i \hat H t/\hbar } =  \sum_{a}   \hat d^\dagger_a  (e^{ i \mathcal{H} t/\hbar})_{ap}, \\
  e^{i \hat H t} \hat d_q  e^{ - i \hat H t /\hbar } =  \sum_{b} (e^{- i \mathcal{H} t/\hbar })_{qb}  \hat d_b .
\end{eqnarray}

If we plug the above equations into Eq. \ref{eq:friction45},  we arrive at
\begin{eqnarray}
\gamma_{\mu \nu} 
&=& - \int_0^\infty dt \:  \sum_{pqab} \partial_{\mu} \mathcal{H}_{pq}    (e^{ i \mathcal{H} t/\hbar})_{ap}  (e^{- i \mathcal{H} t/\hbar })_{qb}    tr_e \left( \hat d^\dagger_a \hat  d_b \partial_{\nu} \hat \rho_{ss}    \right)  \nonumber \\
&=& - \int_0^\infty dt \:  \sum_{pqab} \partial_{\mu} \mathcal{H}_{pq}      (e^{- i \mathcal{H} t/\hbar })_{qb}   \partial_{\nu} \sigma^{ss}_{ba} (e^{ i \mathcal{H} t/\hbar})_{ap}  \nonumber \\
&=& - \int_0^\infty dt \:  Tr_m (  \partial_{\mu}  \mathcal{H} e^{  -i \mathcal{H} t/\hbar } \partial_{\nu} \sigma_{ss}  e^{   i  \mathcal{H}  t/\hbar }  )  .
\end{eqnarray}
Here, we have used the definition of $\sigma^{ss}_{ba} = tr_e \left( \hat d^\dagger_a \hat  d_b\hat \rho_{ss}    \right) $.

The above equation can be recast into the energy domain (with $\eta$ being a positive infinitesimal),
\begin{eqnarray}  
\gamma_{\mu \nu} &=& - \int_0^\infty dt \:  Tr_m (  \partial_{\mu}  \mathcal{H} e^{  -i (\mathcal{H} -i\eta) t/\hbar } \partial_{\nu} \sigma_{ss}  e^{   i  (\mathcal{H}+i\eta)  t/\hbar }  )    \nonumber \\
&=& - \int_0^\infty dt \int_0^\infty dt' \:  Tr_m (  \partial_{\mu}  \mathcal{H} e^{  -i (\mathcal{H} -i\eta) t/\hbar } \partial_{\nu} \sigma_{ss}  e^{   i  (\mathcal{H}+i\eta)  t'/\hbar }  )  \delta (t-t') \nonumber \\
&=&  -  \int_{-\infty}^\infty  \frac{d\epsilon}{2\pi\hbar} \int_0^\infty dt \int_0^\infty dt' \:  Tr_m (  \partial_{\mu}  \mathcal{H} e^{  -i (\mathcal{H} -i\eta) t/\hbar } \partial_{\nu} \sigma_{ss}  e^{   i  (\mathcal{H}+i\eta)  t'/\hbar }  ) e^{i\epsilon(t-t')/\hbar} \nonumber \\
&=&-  \hbar  \int  \frac{d\epsilon}{2\pi} \:  Tr_m \left( \partial_\mu \mathcal{H} \frac{1}{ \epsilon -  \mathcal{H}  +  i\eta} \partial_\nu  \sigma_{ss}  \frac{1}{ \epsilon -  \mathcal{H}  -  i\eta} \right) ,
\end{eqnarray}
which gives us Eq. \ref{eq:frictionSP}.

\section{The lesser self-energy of the total system, $\Pi^<$} \label{app:b}

For the quadratic Hamiltonian with dot-lead separation in Eqs. \ref{eq:ah1}-\ref{eq:ah4},  we imagine embedding the total Hamiltonian (as in Eqs. \ref{eq:ah1}-\ref{eq:ah4}) as the inner Hamiltonian inside yet another, even larger outer bath \footnote{In general, this introduction of the outer bath with infinitesimal  coupling will not affect any system dynamics. There are outstanding issues regarding the effect of the outer bath on entropy production and overall irreversibility, but these quantities will not be calculated here. 
}: 
\begin{eqnarray}
\hat{\tilde{H}}= \hat H_{{inner}} + \hat H_{{outer}}+ \hat V_{{inner-outer}}. 
\end{eqnarray}

Assuming small inner-outer coupling $\hat V_{{inner-outer}}$,  as mediated only through the inner bath (leads), and assuming a completely quadratic Hamiltonian, 
$\Pi^<$ of the inner Hamiltonian can be written as  
\begin{eqnarray} \label{eq:PiLesser}
\Pi^<_{k\alpha, k'\alpha'} = i 2 \eta f_\alpha (\epsilon_{k\alpha} ) \delta_{k,k'} \delta_{\alpha,\alpha'}.
\end{eqnarray}
Here $\eta$ is an positive infinitesimal, which implies that we have added a small dissipation ($\eta$) to all of the non-interacting electrons in the leads (inner bath). $f_\alpha (\epsilon_{k\alpha} )$ is the Fermi function of the $\alpha$ (left or right) lead. 
Clearly,  $\Pi^<$ does not depend on position or energy ($\epsilon'$ as in Eq. \ref{eq:LangrethRule}) if we assume that the inner and outer baths do not depend on $\bold R$.

\subsection{Evaluating $G^<_{mn}$}

As an example of how the definition in the above equation works, we calculate $G^<_{mn} $. 
Starting from the Keldysh equation, $\mathcal{G}^<= \mathcal{G}^R \Pi^< \mathcal{G}^A$, we project the equation onto the dots, 
\begin{eqnarray} \label{eq:g<mn}
G^<_{mn} = \sum_{k\alpha, k'\alpha'} \mathcal{G}^R_{m, k\alpha} \Pi^<_{k\alpha, k'\alpha'} \mathcal{G}^A_{k'\alpha', n} .
\end{eqnarray}
Using the Dyson equation, 
\begin{eqnarray}
\mathcal{G}_{m, k\alpha}   = \sum_{n'} G_{m, n'}   V_{n', k\alpha}  g_{k\alpha} ,\\
\mathcal{G}_{k\alpha, m}   = \sum_{n'}    g_{k\alpha} V_{k\alpha, n'} G_{n', m} , 
\end{eqnarray}
with the zero order retarded and advanced Green's functions for the leads, 
\begin{eqnarray}
g^r_{k\alpha}  = \frac{1}{ \epsilon - \epsilon_{k\alpha}  + i\eta} , g^a_{k\alpha}  = \frac{1}{ \epsilon - \epsilon_{k\alpha}  - i\eta} ,
\end{eqnarray}
we can write Eq. \ref{eq:g<mn} as  
\begin{eqnarray}
G^<_{mn} &=& \sum_{k\alpha, k'\alpha', m', n'} G^R_{m, n'}   V_{n', k\alpha}  g^r_{k\alpha} 2 i \eta f_\alpha(\epsilon_{k\alpha} ) \delta_{k,k'} \delta_{\alpha,\alpha'}  g^a_{k'\alpha'} V_{k'\alpha', m'} G^A_{m', n} \nonumber \\
&=& \sum_{k\alpha, m', n'} G^R_{m, n'}   V_{n', k\alpha}  g^r_{k\alpha} 2 i \eta f_\alpha(\epsilon_{k\alpha} ) g^a_{k\alpha} V_{k \alpha, m'} G^A_{m', n}.
\end{eqnarray}
Note that 
\begin{eqnarray} \label{eq:appg<}
g^r_{k\alpha}  2 i \eta f_\alpha (\epsilon_{k\alpha} ) g^a_{k\alpha}  = \frac{2i\eta}{ (\epsilon - \epsilon_{k\alpha} )^2 + \eta^2}  f_\alpha(\epsilon_{k\alpha} )  = i 2\pi \delta( \epsilon - \epsilon_{k\alpha} ) f_\alpha(\epsilon_{k\alpha} ) = g^<_{k\alpha},
\end{eqnarray}
$g^<_{k\alpha}$ is the zero order lesser Green's functions for the leads. 
Using the standard definition of $\Sigma^<_{n', m'}$
\begin{eqnarray} \label{eq:appSigma<}
\Sigma^<_{n', m'}   = \sum_{k\alpha} V_{n', k\alpha}  g^<_{k\alpha} V_{k \alpha, m'} = \sum_{k\alpha} V_{n', k\alpha}  g^r_{k\alpha} 2 i \eta f_\alpha(\epsilon_{k\alpha} ) g^a_{k\alpha} V_{k \alpha, m'},
\end{eqnarray} 
we arrive at the standard NEGF Langreth equation for $G^<$ for the dots \cite{negf, JH1877notes}:  
\begin{eqnarray} \label{eq:appG<}
G^<_{mn} = \sum_{m', n'} G^R_{m, n'}  \Sigma^<_{n', m'}  G^A_{m', n}.
\end{eqnarray}

\subsection{Evaluating $\mathcal{G}^<_{k\alpha, k'\alpha'}$}

As another example of how to apply the definition in Eq. \ref{eq:PiLesser}, we calculate $\mathcal{G}^<_{k\alpha, k'\alpha'}$ by projecting the Keldysh equation (Eq. \ref{eq:LangrethRule}) onto the leads, 
\begin{eqnarray} \label{eq:appG<kk}
\mathcal{G}^<_{k\alpha, k'\alpha'} = \sum_{k''\alpha''} \mathcal{G}^R_{k\alpha, k''\alpha''} \Pi^<_{k''\alpha'', k''\alpha''} \mathcal{G}^A_{k''\alpha'', k'\alpha'} .
\end{eqnarray}
Again, we have the Dyson equation for the leads: 
\begin{eqnarray} 
 \mathcal{G}^R_{k\alpha, k''\alpha''} &=&  g^r_{k\alpha} \delta_{k\alpha,  k''\alpha''} + \sum_{mn} g^r_{k\alpha} V_{k\alpha, m} G^R_{mn} V_{n, k''\alpha''} g^r_{k''\alpha''} , \\
 \mathcal{G}^A_{k''\alpha'', k'\alpha'} &=&  g^a_{k''\alpha''} \delta_{k''\alpha'',  k'\alpha'} + \sum_{m'n'} g^a_{k''\alpha''} V_{k''\alpha'', m'} G^A_{m'n'} V_{n', k'\alpha'} g^a_{k'\alpha'} . 
\end{eqnarray}
Using Eq. \ref{eq:appg<} ($g^<_{k\alpha} = g^r_{k\alpha} \Pi^<_{k\alpha, k\alpha}  g^a_{k\alpha} $), Eq. \ref{eq:appSigma<} and Eq. \ref{eq:appG<}, we recast $\mathcal{G}^<_{k\alpha, k'\alpha'}$ as
\begin{eqnarray} \label{eq:G<kk}
\mathcal{G}^<_{k\alpha, k'\alpha'} = g^<_{k\alpha} \delta_{k\alpha,  k'\alpha'} &+& \sum_{mn} g^r_{k\alpha} V_{k\alpha, m} G^R_{mn} V_{n, k'\alpha'} g^<_{k'\alpha'}  \nonumber \\
&+&  \sum_{mn} g^<_{k\alpha} V_{k\alpha, m} G^A_{mn} V_{n, k'\alpha'} g^a_{k'\alpha'}  \nonumber \\
&+&  \sum_{mn} g^r_{k\alpha} V_{k\alpha, m} G^<_{mn} V_{n, k'\alpha'} g^a_{k'\alpha'}  
\end{eqnarray}
Recall that $G^R_{mn} = (\epsilon-h-\Sigma^R)^{-1}_{mn}$. Thus we arrive at the standard NEGF result for $\mathcal{G}^<_{k\alpha, k'\alpha'}$.  
Eq. \ref{eq:G<kk} can be derived equivalently by projecting the Dyson equation for the contour-ordered Green's function onto the two different branches of the Keldysh contour, i.e. projecting $ \mathcal{G}^c $ onto $ \mathcal{G}^<$. \cite{negf, JH1877notes}

Lastly, by projecting the Keldysh equation (Eq. \ref{eq:LangrethRule}) onto the appropriate contours in an analogous fashion, we can also derive similar expressions for the dot-lead coupling lesser GF, $\mathcal{G}^<_{m,k\alpha}$.

\end{document}